\documentclass[oneside,a4paper,11pt,shownumbers]{article}
\usepackage{latexsym}
\usepackage{euscript}
\usepackage{epsfig,amsmath,amssymb}

\newcommand{\e}{{\rm e}}
\def\be{\begin{equation}}
\def\ee{\end{equation}}
\def\bea{\begin{eqnarray}}
\def\eea{\end{eqnarray}}

\long\def\symbolfootnote[#1]#2{\begingroup%
\def\thefootnote{\fnsymbol{footnote}}\footnote[#1]{#2}\endgroup} 

 \large\normalsize

\begin{document}

\begin{center}

{\Large \bf The effect of dark strings on semilocal strings}

\vspace*{7mm} {Yves Brihaye $^{a}$
\symbolfootnote[1]{E-mail:yves.brihaye@umh.ac.be} and
Betti Hartmann $^{b}$
\symbolfootnote[2]{E-mail:b.hartmann@jacobs-university.de}}
\vspace*{.25cm}

${}^{a)}${\it Faculte de Sciences, University de Mons, 7000 Mons, Belgium}\\
${}^{b)}${\it School of Engineering and Science, Jacobs University Bremen, 28759 Bremen, Germany}\\

\vspace*{.3cm}
\end{center}

\begin{abstract}
Dark strings have recently been suggested to exist in new models of dark matter
that explain the excessive electronic production in the galaxy.
We study the interaction of these dark strings with semilocal strings
which are solutions of the bosonic sector of the Standard Model in the limit $\sin^2\theta_{\rm w}=1$, where $\theta_{\rm w}$ is the Weinberg angle. While embedded Abelian--Higgs strings
exist for generic values of the coupling constants, we show that
semilocal solutions with non-vanishing condensate inside the string core exist only above a critical value of the  Higgs to gauge boson mass ratio
when interacting with dark strings. Above this critical value, which is greater than unity, the energy per unit length
of the semilocal--dark string solutions is always smaller than that of the embedded Abelian--Higgs--dark string solutions and we show that Abelian--Higgs--dark strings
become unstable above this critical value. Different from the non--interacting case, we would thus expect
semilocal strings to be stable for values of the Higgs to gauge boson mass ratio larger than unity. Moreover, the one-parameter family of solutions present in the non-interacting case
ceases to exist when semilocal strings interact with dark strings.

\end{abstract}

\section{Introduction}
There is strong observational evidence \cite{observation_dm} that approximately $22\%$ of the total energy density of the universe is in the form of dark matter. Up until now it is unclear what this dark matter should be made of. One of the favourite candidates are Weakly Interacting Massive Particles (WIMPs) which
arise in extensions of the Standard Model.
Recently, new theoretical models of the dark matter sector have been proposed \cite{dark}, in which
the Standard Model is coupled to the dark sector via an attractive interaction term.
These models have been motivated by new astrophysical observations \cite{observation_ee} which show an excess 
in electronic production in the galaxy. Depending on the experiment, the energy of these
excess electrons is between a few GeV and a few TeV.
One possible explanation for these observations is the annihilation of dark matter 
into electrons. Below the GeV scale, the interaction term in these models is basically of the form of a direct
coupling between the U(1) field strength tensor of the dark matter sector and the U(1) field strength tensor of
electromagnetism. The U(1) symmetry of the dark sector  has to be spontaneously broken, otherwise a ``dark photon'' background leading to observable consequences would exist.

Consequently, it has been shown that the dark sector can have string-like solutions, denominated ``dark strings'' 
and the observational consequences of the interaction of these dark strings with the Standard Model
have been discussed \cite{vachaspati}.

Topological defects are believed to have formed in the numerous phase transitions in the early
universe due to the Kibble mechanism \cite{topological_defects}.
While magnetic monopoles and domain walls, which result from the spontaneous
symmetry breaking of a spherical and parity symmetry, respectively, 
are catastrophic for the universe since they would overclose it, cosmic strings
are an acceptable remnant from the early universe. These objects 
form whenever an axial symmetry gets spontaneously broken and (due to topological arguments)
are either infinitely long or exist in the form of cosmic string loops. Numerical
simulations of the evolution of cosmic string networks have shown that
these reach a scaling solution, i.e. their contribution to the total energy density
of the universe becomes constant at some stage. The main mechanism that allows
cosmic string networks to reach this scaling solution is the formation
of cosmic string loops due to self-intersection and the consequent decay of these loops
under the emission of gravitational radiation.

For some time, cosmic strings were believed to be responsible for the structure
formation in the universe. New Cosmic Microwave background (CMB) data, however, clearly
shows that the theoretical power spectrum associated to Cosmic strings
is in stark contrast to the observed power spectrum. However, there has been
a recent revival of cosmic strings since it is now believed that cosmic  strings
might be linked to the fundamental strings of string theory \cite{polchinski}.

While perturbative fundamental strings were excluded to be observable on cosmic scales
for many reasons \cite{witten}, there are now new theories containing
extra dimensions, so-called brane world model, that allow to lower the fundamental
Planck scale down to the TeV scale. This and the observation that
cosmic strings generically form at the end of inflation in inflationary models
resulting from String Theory \cite{braneinflation} and Supersymmetric Grand Unified Theories \cite{susyguts}
has boosted the interest in comic string solutions again.

Different field theoretical models describing cosmic strings have been investigated.
The $U(1)$ Abelian--Higgs model possesses string--like solutions \cite{no}. This is a simple toy
model that is frequently used to describe cosmic strings. However, the symmetry breaking
pattern $U(1)\rightarrow 1$ has very likely never occurred in the evolution of the universe.
Consequently, more realistic models with gauge group $SU(2)\times U(1)$ and symmetry
 breaking $SU(2)\times U(1)\rightarrow U(1)$ have been considered and it has been
shown that these models have string--like solutions \cite{semilocal,gors}. Semilocal strings
are solutions of a $SU(2)_{global}\times U(1)_{local}$ model which -- in fact --
corresponds to the Standard Model of Particle physics in the limit $\sin^2\theta_{\rm w}=1$, where
$\theta_{\rm w}$ is the Weinberg angle. The simplest semilocal string solution is an
embedded Abelian--Higgs solution \cite{semilocal}. A detailed analysis of the stability
of these embedded solutions has shown \cite{hindmarsh} that they are unstable (stable)
if the Higgs boson mass is larger (smaller) than the gauge boson mass. In the case
of equality of the two masses, the solutions fulfill a Bogomolny--Prasad--Sommerfield (BPS) \cite{bogo}
bound such that their energy per unit length is directly proportional to the winding number.
Interestingly, it has been observed \cite{hindmarsh} that in this BPS limit, a one-parameter
family of solutions exists: the Goldstone field can form a non-vanishing condensate
inside the string core and the energy per unit length is independent of the value of this condensate.
These solutions are also sometimes denominated ``skyrmions'' and have been related to the zero-mode present
in the BPS limit.

In this paper, we consider the interaction of dark strings with string--like solutions
of the Standard Model in the specific limit $\sin^2\theta_{\rm W}=1$. The two sectors
interact via an attractive interaction that couples the two $U(1)$ field strength tensors to each other.
This type of interaction has been studied before in \cite{ha}, where the interaction
between Abelian--Higgs strings and dark strings has been investigated. It has been 
found that a BPS bound exists that depends on the interaction paramater and that Abelian--Higgs strings
and dark strings can form bound states.

Our paper is organized as follows: in Section 2, we give the model, the equations of motion, the boundary
conditions and the asymptotics. In Section 3, we present our numerical results and Section 4 contains
our conclusions.

\section{The model}
We study the interaction of a $SU(2)_{global}\times U(1)_{local}$ model, which
has semilocal strings solutions \cite{semilocal}
with the low energy dark sector, which is a $U(1)$ Abelian-Higgs model.

The matter Lagrangian
${\cal L}_{m}$ reads:
\begin{equation}
{\cal L}_{m}=(D_{\mu} \Phi)^{\dagger} D^{\mu} \Phi-\frac{1}{4} F_{\mu\nu} F^{\mu\nu}
-\frac{\lambda_1}{2}(\Phi^{\dagger}\Phi-\eta_1^2)^2
+(D_{\mu} \xi)^* D^{\mu} \xi-\frac{1}{4} H_{\mu\nu} H^{\mu\nu}
- \frac{\lambda_2}{2}(\xi^*\xi-\eta_2^2)^2  + \frac{\varepsilon}{2} F_{\mu\nu}H^{\mu\nu}
\end{equation} 
with the covariant derivatives $D_\mu\Phi=\nabla_{\mu}\Phi-ie_1 A_{\mu}\Phi$,
$D_\mu\xi=\nabla_{\mu}\xi-ie_2 a_{\mu}\xi$
and the
field strength tensors $F_{\mu\nu}=\partial_\mu A_\nu-\partial_\nu A_\mu$, 
$H_{\mu\nu}=\partial_\mu a_\nu-\partial_\nu a_\mu$  of the two U(1) gauge potential $A_{\mu}$, $a_{\mu}$ with coupling constants $e_1$
and $e_2$.
$\Phi=(\phi_1,\phi_2)^T$ is a complex scalar doublet, while $\xi$ is a complex scalar field.
The gauge fields have masses $M_{W,i}=\sqrt{2}e_i \eta_i$, $i=1,2$, while the Higgs fields
have masses $M_{H,i}=\sqrt{2\lambda_i} \eta_i$, $i=1,2$.
The term proportional to $\varepsilon$ is the interaction term \cite{vachaspati}.
To be compatible with observations, $\varepsilon$ should be on the order of $10^{-3}$. 

\subsection{The Ansatz}

For the matter and gauge fields, we have \cite{semilocal,hindmarsh,no}:
\begin{equation}
\phi_1(\rho,\varphi)=\eta_1 h_1(\rho)e^{i n\varphi} \ \ , \ \  \phi_2(\rho)=\eta_1 h_2(\rho) \ \ , \ \
\xi(\rho,\varphi)=\eta_2 f(\rho)e^{i m\varphi} 
\end{equation}
\begin{equation}
A_{\mu}dx^{\mu}=\frac {1}{e_1}(n-P(\rho)) d\varphi \ \ , \ \ a_{\mu}dx^{\mu}=\frac {1}{e_2}(m-R(\rho)) d\varphi \ .
\end{equation}
$n$ and $m$ are integers indexing the vorticity of the two Higgs fields  around the $z-$axis.
In the following, we will refer to solutions with $h_2(\rho)\equiv 0$ as ``embedded Abelian--Higgs solutions'', while solutions with $h_2(\rho)\neq 0$ will be referred to as ``semilocal solutions''.
Note that in the case $\varepsilon=0$, the solutions of the semilocal sector of
our model are often also denominated ``skyrmions''.

\subsection{Equations of motion}
We define the following dimensionless variable $x=e_1\eta_1 \rho$, which measures the radial
distance in units of $M_{W,1}/\sqrt{2}$. 

Then, the total Lagrangian ${\cal L}_m \rightarrow {\cal L}_m/(\eta_1^4 e_1^2)$ depends only on the following dimensionless coupling constants

\begin{equation}
\beta_i=\frac{\lambda_i}{e_1^2}=\frac{M^2_{H,i}}{M^2_{W,1}}\frac{\eta_1^2}{\eta_i^2} \ , \ i=1,2   \ , \ \  g=\frac{e_2}{e_1} \ \ , \  \ \
q=\frac{\eta_2}{\eta_1} \ \ . \ \ 
\end{equation}
 Varying the action with respect to the matter fields we
obtain a system of five non-linear differential equations. The Euler-Lagrange equations for the matter field functions read:
\begin{equation}
\label{eq1}
(xh_1')'=\frac{P^2 h_1}{x}+\beta_1x(h_1^2+h_2^2-1)h_1
\end{equation}
\begin{equation}
\label{eq2}
(xh_2')'=\frac{(n-P)^2 h_2}{x}+\beta_1x(h_1^2+h_2^2-1)h_2
\end{equation}
\begin{equation}
\label{eq3}
(xf')'=\frac{R^2f}{x}+\beta_2x(f^2-q^2)f
\end{equation}
\begin{equation}
\label{eq4}
(1-\varepsilon^2)\left(\frac{P'}{x}\right)'=2 \frac{h_1^2 P}{x} -2\frac{(n-P)h_2^2}{x} + 2\varepsilon g \frac{R f^2}{x} \ ,
\end{equation}
\begin{equation}
\label{eq5}
(1-\varepsilon^2)\left(\frac{R'}{x}\right)'=2 g^2 \frac{f^2 R}{x} + 2\varepsilon g \left(\frac{P h_1^2}{x}-\frac{(n-P)h_2^2}{x}\right) \ ,
\end{equation}
where the prime now and in the following denotes the derivative with respect to $x$.

\subsection{Energy per unit length and magnetic fields}

The non-vanishing components of the energy-momentum tensor are (we use the notation
of \cite{clv}):
\begin{eqnarray}
T_0^0 &=& e_s + e_v + e_w + u  \ \ , \ \ 
T_x^x = -e_s - e_v + e_w + u \nonumber \\
T_{\varphi}^{\varphi} &=&e_s - e_v - e_w + u \ \ , \ \ T_z^z =  T_0^0 
\end{eqnarray}
where
\begin{equation}
\label{contributions}
e_s= (h_1')^2 + (h_2')^2 + (f')^2   \ \ \ , \ \ \ e_v = \frac{(P')^2}{2 x^2} + \frac{(R')^2}{2 g^2 x^2} -\frac{\varepsilon}{g}\frac{R'P'}{x^2} \ \ \ , \ \ \ e_w = \frac{h_1^2 P^2}{x^2} + \frac{h_2^2 (n-P)^2}{x^2} + \frac{R^2 f^2}{x^2} \end{equation}
and
\begin{eqnarray}
u & = & \frac{\beta_1}{2}\left(h_1^2+h_2^2-1\right)^2 + \frac{\beta_2}{2} \left(f^2-q^2\right)^2  \ . 
\end{eqnarray}

We define as inertial energy per unit length  of a solution describing the interaction of a semilocal
string with winding $n$ and a dark string with winding $m$:
\begin{equation}
 \mu^{(n,m)}=\int \sqrt{-g_3} T^0_0 dx d\varphi
\end{equation}
where $g_3$ is the determinant of the $2+1$-dimensional space-time given by $(t,x,\varphi)$.
This then reads:
\begin{equation}
 \mu^{(n,m)}=2\pi\int_{0}^{\infty} x \left(\varepsilon_s + \varepsilon_v + \varepsilon_w + u\right) \ dx
\end{equation}
Note that the string tension $T=\int \sqrt{-g_3} \ T^z_z dx d\varphi$ is equal to the energy per unit length. There are a few special case, in which energy bounds can be given:
\begin{enumerate}
 \item
For $h_2(x)\equiv 0$, the energy per unit length of the solution is
given by:
\begin{equation}
\mu^{(n,m)}=2\pi n \eta_1^2 g_1(\beta_1) + 2\pi m \eta_1^2   g_2(\beta_2) 
\end{equation}
where $g_1$ and $g_2$ are functions that depend only weakly on $\beta_1$ and $\beta_2$, respectively.
The energy bound is fulfilled, when the functions $g_1$ and $g_2$ become equal to unity.
This happens at $\beta_1=\beta_2=1/(1-\varepsilon)$  and $n=m$ \cite{ha}.

\item For $\varepsilon=0$ and $h_2(x)\neq  0$, the energy per unit length of the solution is
given by
\begin{equation}
\mu^{(n,m)}=2\pi n \eta_1^2 + 2\pi m \eta_2^2   g_2(\beta_2) 
\end{equation}
where $g_2$ is a  function that depends only weakly on $\beta_2$ with $g_2(1)=1$.
Note that the solution of the semilocal sector exists only for $\beta_1=1$ and fulfills the BPS bound
for all choices of $h_2(0)$.

\end{enumerate}

The magnetic fields associated to the solutions are given by \cite{ha}~:
\begin{equation}
\label{magnetic}
 B_z(x)=\frac{-P'(x)+\frac{\varepsilon}{g} R'(x)}{e_1 x} \ \ \ {\rm and} \ \ \   b_z(x)=-\sqrt{1-\varepsilon^2}\frac{R'(x)}{e_2 x}  \ ,  
\end{equation}
respectively, where we have used the fact that the part of the Lagrangian containing
the field strength tensors can be rewritten as \cite{vachaspati}~:
\begin{equation}
 -\frac{1}{4} F_{\mu\nu} F^{\mu\nu}-\frac{1}{4} H_{\mu\nu} H^{\mu\nu}+ \frac{\varepsilon}{2} F_{\mu\nu}H^{\mu\nu} \Rightarrow -\frac{1}{4} G_{\mu\nu} G^{\mu\nu} -\frac{1}{4}(1-\varepsilon^2) H_{\mu\nu} H^{\mu\nu}
\end{equation}
with $G_{\mu\nu}=\partial_{\mu} \tilde{A}_{\nu}- \partial_{\nu} \tilde{A}_{\mu}$
where $\tilde{A}_{\mu}=A_{\mu}-\varepsilon a_{\mu}$.
The corresponding magnetic fluxes $\int d^2x \ B$ are
\begin{equation}
 \Psi= \frac{2\pi}{e_1}\left(n-\frac{\varepsilon}{g} m\right) \ \ {\rm and} \ \ 
\psi=\sqrt{1-\varepsilon^2} \ \frac{2\pi m}{e_2} \ ,
\end{equation}
respectively. Obviously, these magnetic fluxes are not quantized for generic $\varepsilon$.

\subsection{Boundary conditions and asymptotics}
The requirement of regularity at the origin leads to the  following boundary 
conditions:
\begin{equation}
h_1(0)=0 \ , \ h_2'(0)=0 \ ,  \ f(0)=0 \ , \ P(0)=n \ , \ R(0)=m
\label{bc1}
\end{equation}
For $h_2(0)=0$, the semilocal strings correspond to embedded Abelian--Higgs 
strings. Here, we are mainly interested in constructing solutions that are truly semilocal, i.e. we require $h_2(0)\neq 0$.
The finiteness of the energy per unit length requires:
\begin{equation}
h_1(\infty)=1 \ , \ h_2(\infty)=0 \ ,  \ f(\infty)=q \ , \ P(\infty)=0 \ , \ R(\infty)=0  \ .
\end{equation}

The asymptotic behaviour for $x\rightarrow \infty$ depends crucially on whether the function $h_2(x)\equiv 0$ or
$h_2(x)\neq  0$. 
\begin{enumerate}
 \item For $h_2(x)\equiv 0$ we find:
\begin{eqnarray}
 P (x\rightarrow \infty) &=& - \sqrt{x} \  m_{12} \ \left[C_1  \exp\left(-x\beta_+\right) + C_2  \exp\left(-x \beta_-\right) \right] + .... \\
 R(x\rightarrow \infty) &=& \sqrt{x} \ 
 \left[C_1 \ m_{11}(\beta_+) \exp\left(-x\beta_+\right) + C_2 \ m_{11}(\beta_-) \exp\left(-x \beta_-\right) \right]+ ...
\end{eqnarray}
where $C_1$ and $C_2$ are constants,  $m_{11}(\beta_{\pm})= (1-\varepsilon^2)\beta_{\pm}^2 - 2$ and $m_{12} = -2 \varepsilon q^2 g$. The $\beta_{\pm}$ are positive and are given by
\begin{equation}
   \beta^2_{\pm} = \frac{1+q^2 g^2 \pm \sqrt{(1-q^2 g^2)+ 4 \varepsilon q^2 g^2}}{1-\varepsilon^2}
\end{equation}
The numerical evaluation (see below) shows that for specific values of the coupling constants the constants $C_1$ and $C_2$
have opposite sign. Hence, the function $R(x)$ can possess a node asymptotically which we have confirmed
numerically. However, the numerics has shown that these type of solutions exist only
for values of $\varepsilon$ of order one. Hence, we don't present them here since we believe that
they are unphysical.

For the scalar fields, we find 
\begin{eqnarray}
h_1 (x\rightarrow \infty) &=& 1 + \frac{C_3}{\sqrt{x}} \exp\left(-x\sqrt{2\beta_1}\right) 
+ \frac{c_+}{x} \exp\left(-2x \beta_+\right) + \frac{c_-}{x} \exp\left(-2x \beta_-\right) + ...\\
f(x\rightarrow \infty) &=& q + \frac{C_4}{\sqrt{x}} \exp\left(-x\sqrt{2\beta_2}\right) 
+ \frac{d_+}{x} \exp\left(-2x \beta_+\right) + \frac{d_-}{x} \exp\left(-2x \beta_-\right) + ...
\end{eqnarray}
$C_3$ and $C_4$ are two constants, while $c_{\pm}$, $d_{\pm}$ depend on 
the constants $C_1, \dots, C_4$ and on $\beta_1$ and $\beta_2$.

\item For $h_2(x)\neq 0$ we find~:
\begin{equation}
\label{as_gauge}
P(x\rightarrow \infty)= \frac{n c^2}{x^{2n}} + ... \ \ , \ \ R(x\rightarrow \infty)= \frac{c_R}{x^{2n+2}} + ...
 \end{equation}
for the gauge field functions. Here $c$, $c_R$ are constants that depend on the values of the coupling constants.
For the scalar and Higgs field functions we have
\begin{equation}
\label{as_scalar}
h_1(x\rightarrow \infty) = 1 - \frac{c^2}{2} \frac{1}{x^{2n}} + ... \ \ , \ \ h_2(x) = \frac{c}{x^n} + ... \ \ , \ \ 
f(x\rightarrow \infty)= q - \frac{c_R^2}{2 q \beta_2} \frac{1}{x^{4n+6}} + ....
\end{equation}

\end{enumerate}
Obviously, the presence of the scalar field $h_2(x)$ changes the asymptotics drastically.
While for $h_2(x)\equiv 0$, the gauge and Higgs fields decay exponentially, they have power-law 
decay for $h_2(x)\neq 0$. 

\subsection{Stability}

Following the investigation in the case $\varepsilon=0$ \cite{hindmarsh}, we are interested
in the stability of the embedded Abelian--Higgs string coupled to a dark string.
In order to do that we will study
the normal mode along a very specific (but standard) direction
in perturbation space about the embedded Abelian--Higgs string coupled to a dark string. We consider the perturbation
\begin{equation}
      h_1(x)=\tilde{h}_1(x) \ , \ h_2(x)= \e^{i\omega t}\eta(x) \ , \ 
      P(x) = \tilde{P}(x) \ , \ R(x) = \tilde{R}(x) \ , \ f(x) = \tilde{f}(x)
\end{equation}
where the tilded functions denote the profiles of an embedded Abelian--Higgs string
coupled to a dark string, i.e. solutions to the equations (\ref{eq1}),(\ref{eq3}), (\ref{eq4}) and (\ref{eq5}) for $h_2(x)\equiv 0$. The perturbation is denoted by $\eta$
and the parameter $\omega$ is real. 
Inserting this perturbation into (\ref{eq2}) and keeping only the linear terms in $\eta$ leads to the
 linear eigenvalue equation~:
\be
\label{sta}
       \left(-\frac{d^2}{dx^2} - \frac{1}{x} \frac{d}{dx} + V_{eff}\right)\eta(x) =  \omega^2 \eta(x) \ \ , 
       \ \ V_{eff} = \frac{(n-\tilde{P}(x))^2}{x^2} + \beta_1 (\tilde{h}_1(x)^2-1)
\ee
The spectrum of the linear operator entering in (\ref{sta}) consists of a continuum for $\omega^2 > 0$ 
and of a finite number of  bound states (or normalisable solutions) for $\omega^2 < 0$.
In the latter case, the solutions fulfill
\be
     \eta(0) = 1 \ , \ \eta'(0) = 0 \ \ \ {\rm with} \ \  \ \eta(x) \to e^{-|\omega| x} \ {\rm for} \ x \to \infty
\ee 
where we have fixed  the arbitrary normalisation by choosing $\eta(0)=1$.

Only bound states are of interest to us since they signal the presence of an instability. 
  It should be pointed out that the functions $\tilde{P}(x)$, $\tilde{h}_1(x)$ entering in the
 effective potential feel the effect of the
dark sector since the corresponding equations are directly coupled.

\section{Numerical results}
For all our numerical calculations, we have chosen $q=g=1$.

\subsection{Stability of the embedded Abelian--Higgs--dark strings}
\begin{figure}[!htb]
\centering
\leavevmode\epsfxsize=12.0cm
\epsfbox{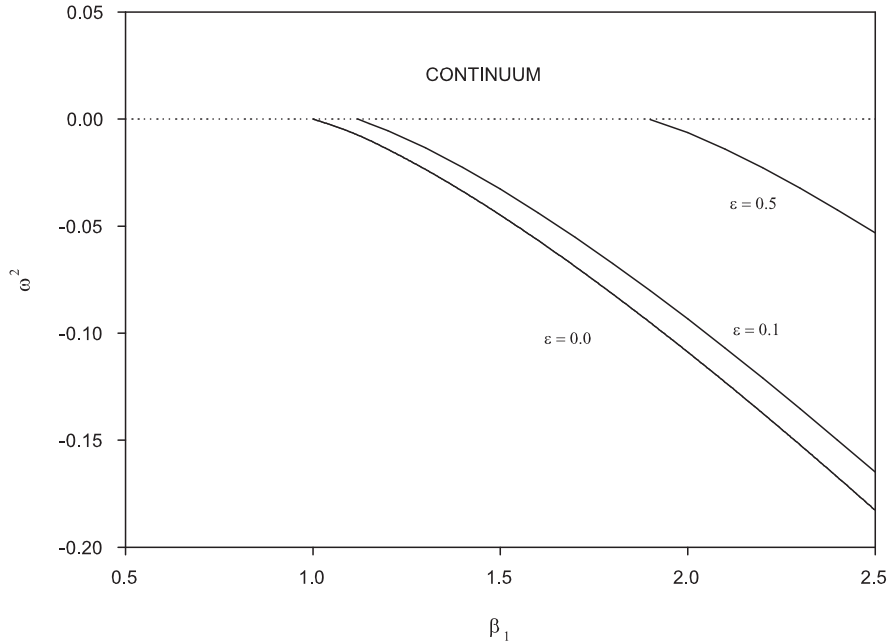}\\
\caption{\label{fignew} We give the value of $\omega^2$ (see (\ref{sta}) in dependence
on $\beta_1$ for three different choices of $\varepsilon$ including the non--interacting
case $\varepsilon=0$. Here $n=m=1$ and $\beta_2=1$. }
\end{figure}

We have first studied the stability of the embedded Abelian--Higgs strings coupled
to dark strings by investigating the bound states of (\ref{sta}) for different
values of $\varepsilon$. Our results for $n=m=1$ and $\beta_2=1$ are shown in Fig.\ref{fignew}.

For $\varepsilon=0$ we recover the result of \cite{hindmarsh} that the
embedded--Abelian-Higgs strings are unstable for $\beta_1 >1$. 
For $\varepsilon \neq 0$, we observe that the larger $\varepsilon$, the larger the
ratio of Higgs to gauge boson mass $\beta_1$ at which the embedded Abelian--Higgs strings
coupled to dark strings become unstable. In the following, we will denote the value of $\beta_1$ at which
$\omega^2=0$ $\beta_1^{cr}$.  
With view to the observations for the $\varepsilon=0$ case, we would thus expect additional
solutions with $h_2(x)\neq 0$ for $\beta_1 > \beta_1^{cr}$. 
In section 3.2, we will discuss the properties of these solutions.
 
Let us also remark that our analysis does not reveal the occurence of additional unstable modes in the sector explored.

\subsection{Properties of semilocal--dark strings}
In the case $\varepsilon=0$, the two sector do not interact and for the semilocal sector
two different types of solutions are possible: (a) embedded Abelian--Higgs solutions with
$h_2(x)\equiv 0$ which exist for generic choices of $\beta_1$ \cite{semilocal} and 
(b) semilocal strings (``skyrmions'') with $h_2(x)\neq 0$ which exist only for $\beta_1=1$ \cite{hindmarsh}.
In the latter case, it was shown that 
there is a zero mode associated to the fact that the energy of the ``skyrmions''
does not depend on the value of $h_2(0)$. 

The case with $\varepsilon\neq 0$ and $h_2(x)\equiv 0$ corresponds hence to the case
of an embedded Abelian--Higgs string interacting with a dark string. The equations
of motion that describe this case are exactly those studied in \cite{ha}.
In \cite{ha}, the interaction of a dark string with an Abelian--Higgs string has been studied
in detail. Since the only difference between an Abelian--Higgs string and
an embedded Abelian--Higgs string is the stability  -- see section 3.1 -- we do not discuss this case in detail in this paper and focus on the
case of semilocal strings interacting with dark strings. We have solved the
differential equations subject to the boundary conditions numerically using the
ODE solver COLSYS \cite{colsys}.

\begin{figure}[!htb]
\centering
\leavevmode\epsfxsize=12.0cm
\epsfbox{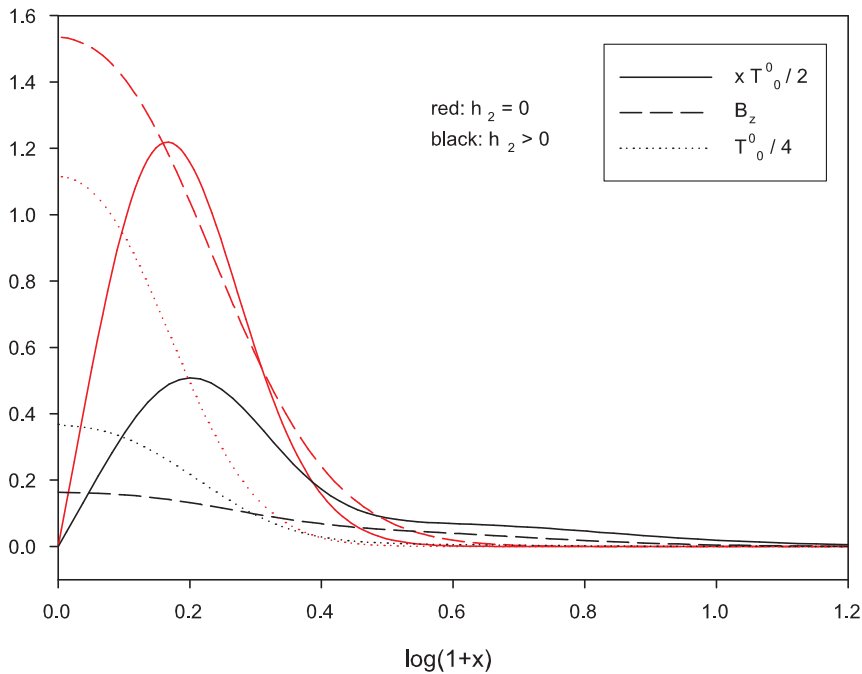}\\
\caption{\label{fig0} We give the profiles of the energy density $T_0^0$, the effective energy density $x\cdot T^0_0$ as well as the
magnetic field $B_z$ (see (\ref{magnetic})) for
$\varepsilon=1/6$, $\beta_1=3$ and $\beta_2=(1-\varepsilon)^{-1}=1.2$. We compare semilocal--dark string
solutions with $h_2(0) > 0$ (black) and embedded Abelian--Higgs--dark string solutions with $h_2(0)=0$ (red).}
\end{figure}

To see the difference between embedded Abelian--Higgs--dark string solutions
and semilocal--dark string solutions, we present the energy density $T_0^0$, the effective
energy density $xT_0^0$ as well as the magnetic field $B_z$ (see (\ref{magnetic}))
in Fig.\ref{fig0} for $\varepsilon=1/6$, $\beta_1=3$ and $\beta_2=(1-\varepsilon)^{-1}=1.2$. Clearly, the effective energy density tends to zero very quickly for the
embedded--Abelian--Higgs--dark string, while for the semilocal--dark string is has a long tail which
results from the power--law fall off of the functions. Moreover, the magnetic field $B_z$
tends to zero exponentially for the embedded Abelian--Higgs--dark strings, while it falls
off power--like for the semilocal--dark strings. Hence, the core of the
magnetic flux tube of the latter solution is not well defined.

While for $\varepsilon=0$ solutions with $h_2(x)\neq 0$ exist only for
$\beta_1=1$, the situation is different here.
For $\varepsilon\neq 0$, we find solutions for generic values of $\beta_1$, i.e. different from
unity. In fact, the solutions exist only for $\beta_1$ larger than a critical 
value, $\beta_1^{cr}$,  which depends on the choice of the winding numbers
and other coupling constants, in particular $\varepsilon$. Moreover, we observe that
the $\beta_1$ at which semilocal--dark strings exist is a function of $h_2(0)$.
While for $\varepsilon=0$, $\beta_1=1$ for all choices of $h_2(0)$, we find that for $\varepsilon\neq 0$
the choice of $h_2(0)$ fixes the value of $\beta_1$.

\begin{figure}[!htb]
\centering
\leavevmode\epsfxsize=12.0cm
\epsfbox{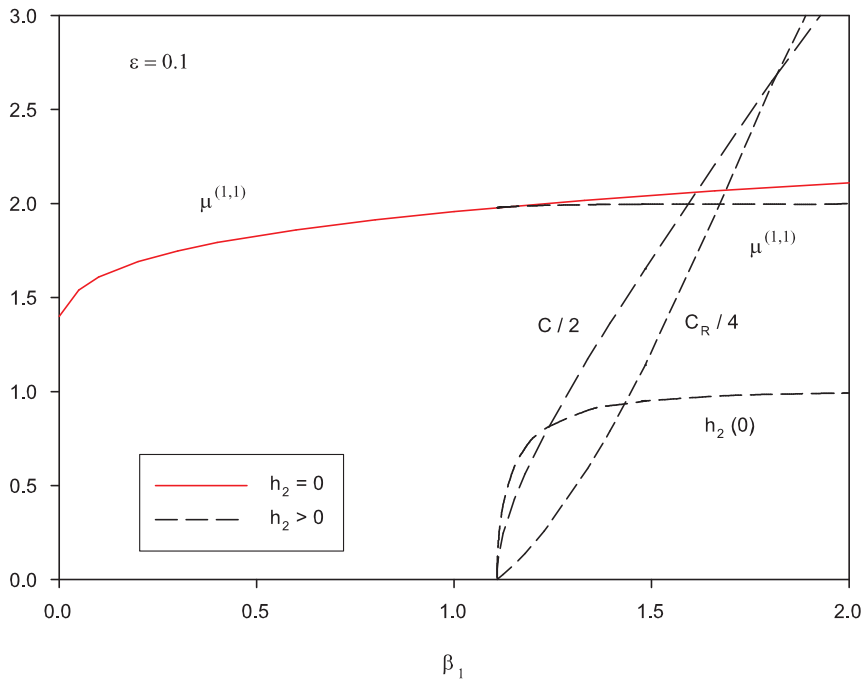}\\
\caption{\label{fig1} The energy per unit length $\mu^{(1,1)}$ (in units of $2\pi\eta_1^2$) as well as the value of $h_2(0)$ and the asymptotic constants  $c$ and $c_R$ (see (\ref{as_gauge}), 
(\ref{as_scalar})) of the semilocal--dark  string solutions are shown in dependence on $\beta_1$ for $\varepsilon=0.1$, $\beta_2=1$ and $n=m=1$ (dashed).
For comparison, we also give the energy per unit length of the embedded Abelian--Higgs--dark string solution (solid).}
\end{figure}

\begin{figure}[!htb]
\centering
\leavevmode\epsfxsize=12.0cm
\epsfbox{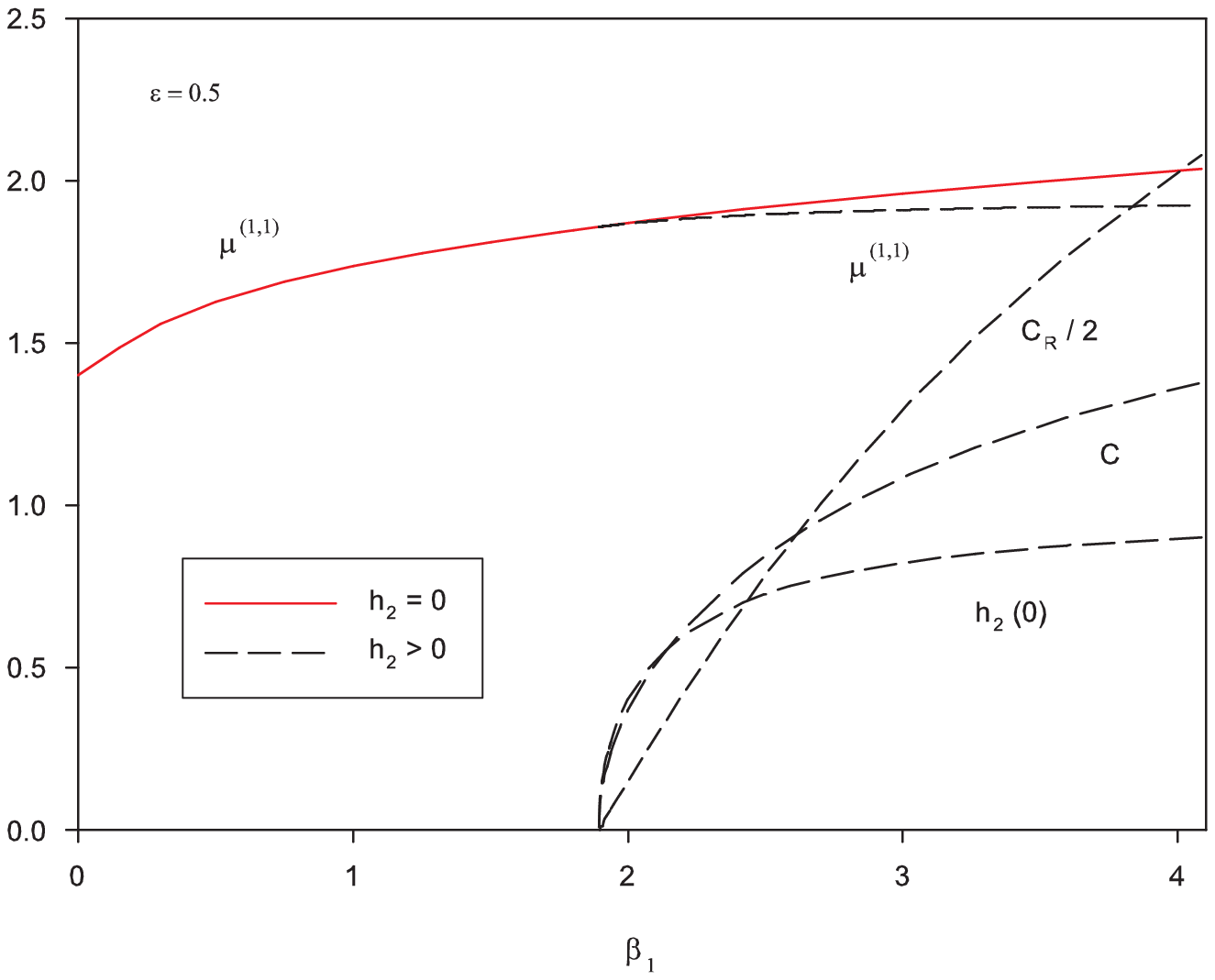}\\
\caption{\label{fig2}The energy per unit length $\mu^{(1,1)}$ (in units of $2\pi\eta_1^2$) as well as the value of $h_2(0)$ and the asymptotic constants  $c$ and $c_R$ (see (\ref{as_gauge}), (\ref{as_scalar})) of the semilocal--dark string solution are shown in dependence on $\beta_1$ for $\varepsilon=0.5$, $\beta_2=1$ and $n=m=1$ (dashed). For comparison, we also give the energy per unit length of the embedded Abelian--Higgs--dark string solution (solid). }
\end{figure}

At $\beta_1^{cr}$ the branch of solutions describing a semilocal string in
interaction with a dark string bifurcates with the branch of solutions describing
the interaction of an embedded Abelian--Higgs string with a dark string. 
This is shown in Fig.s \ref{fig1},\ref{fig2} for $\varepsilon=0.1$ and $\varepsilon=0.5$, respectively. Note that $\beta_1^{cr}$ is exactly the value at which the
embedded Abelian--Higgs--dark strings become unstable.

Here, we give the value of $h_2(0)$ in dependence on $\beta_1$ for $n=m=1$ and $\beta_2=1.0$.
Clearly at some $\beta_1^{cr}$, $h_2(0)$ tends to zero which means that $h_2(x)\equiv 0$.
Here the semilocal--dark string solutions bifurcate with the embedded Abelian--Higgs--dark string solutions.
We also compare the energy per unit length of the two types of solutions. Clearly,
whenever semilocal--dark string solutions exist, they have lower energy than 
the corresponding embedded Abelian--Higgs--dark string solutions. Moreover, the larger $\beta_1$, the bigger
is the difference between the two energies per unit length.
We would thus expect the semilocal solutions to be stable with respect to the decay into the
embedded Abelian solutions when coupled to dark strings. 
We also present the values of the asymptotic constants $c$ and $c_R$ (see (\ref{as_gauge}), (\ref{as_scalar})). These vanish identically at $\beta_1=\beta_1^{cr}$.

\begin{figure}[!htb]
\centering
\leavevmode\epsfxsize=14.0cm
\epsfbox{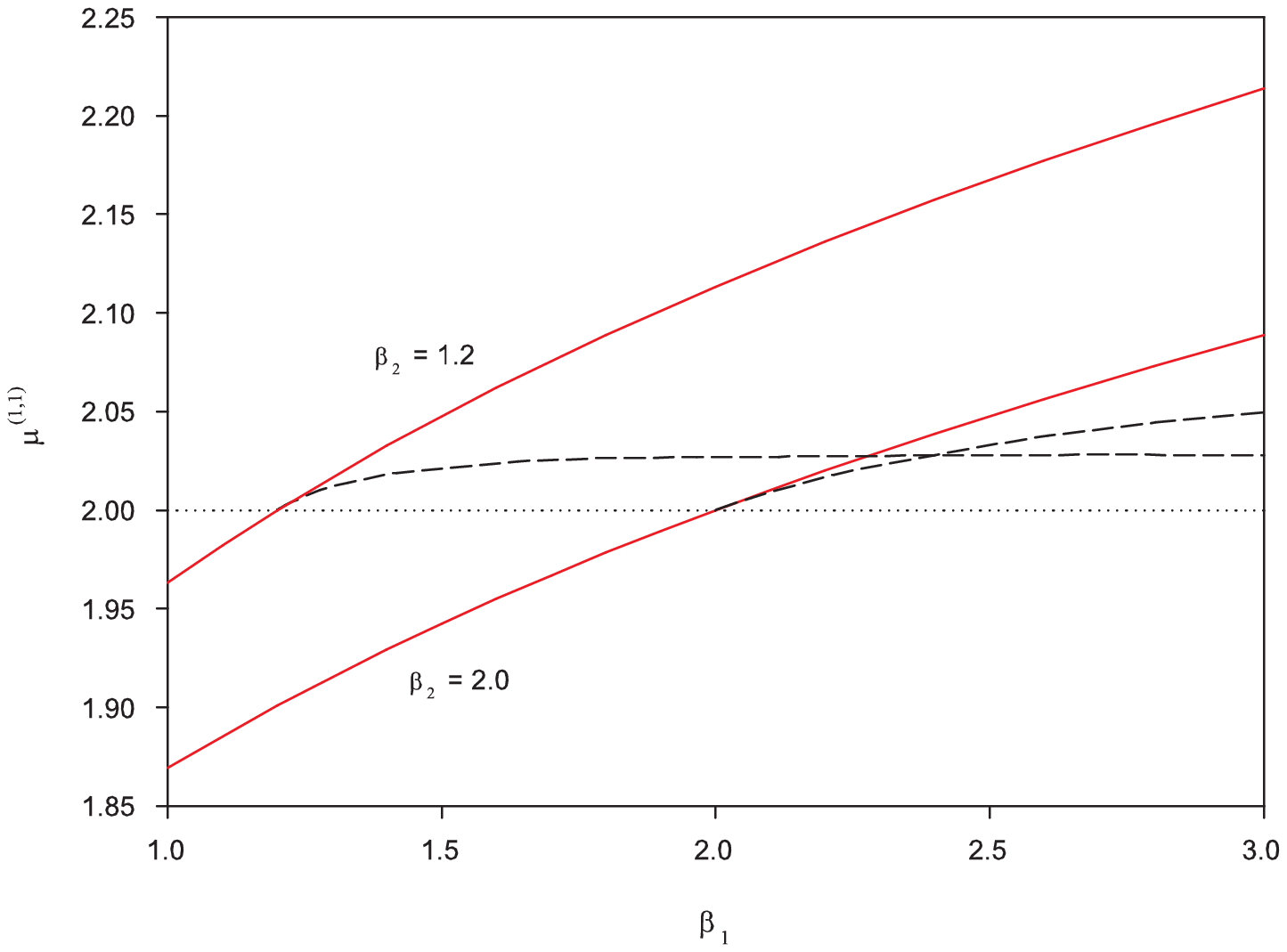}\\
\caption{\label{fig4} The energy per unit length $\mu^{(1,1)}$ (in units
of $2\pi\eta_1^2$) is shown for semilocal strings interacting with dark strings as function of $\beta_1$ for $\beta_2=(1-\varepsilon)^{-1}$ with $\varepsilon=0.5$ and $\varepsilon=1/6$, respectively (dashed). 
For comparison, we also give the energy per unit length of the corresponding embedded Abelian--Higgs
solutions interacting with dark strings (solid).}
\end{figure}

\begin{figure}[!htb]
\centering
\leavevmode\epsfxsize=14.0cm
\epsfbox{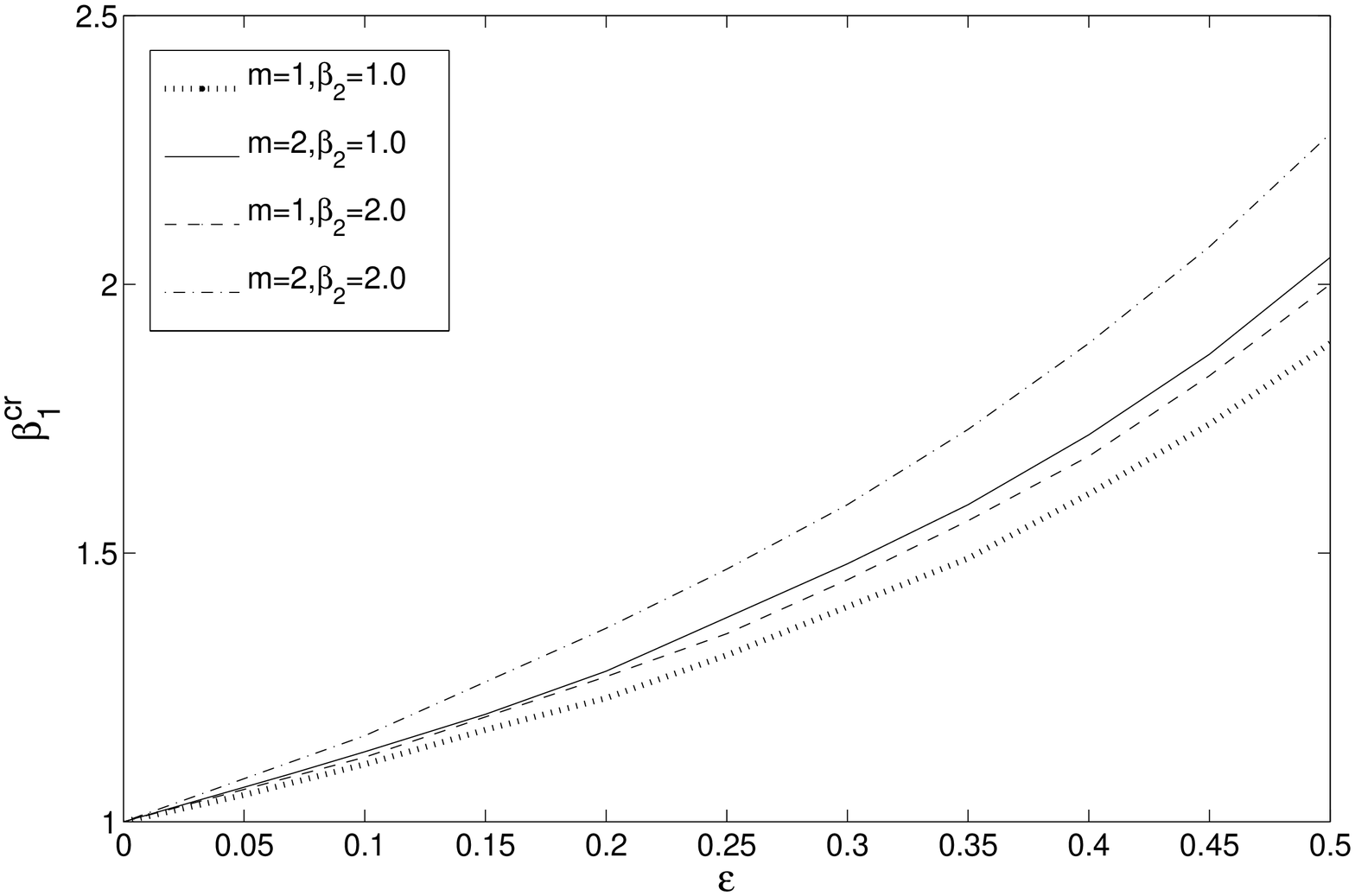}\\
\caption{\label{fig3} The value of $\beta_1^{cr}$ at which the branch of semilocal
solutions bifurcates with the branch of embedded Abelian--Higgs solutions  is shown as function
of $\varepsilon$ for $m=1$, $m=2$, respectively and $\beta_2=1.0$, $\beta_2=2.0$, respectively.  }
\end{figure}

In general, $\beta_1^{cr}$ will depend on the choice of $\beta_2$, $n$ and $m$: $\beta_1^{cr}(\beta_2,n,m)$.
As shown in \cite{ha} in the limit $h_2(x)\equiv 0$ a BPS bound exists for $\beta_1=\beta_2=(1-\varepsilon)^{-1}$ and $n=m$. In this limit, the energy per unit length (in units
of $2\pi\eta_1^2$) is just $n+m=2n$. We have studied the dependence of the energy per unit length
on $\beta_1$ for $\beta_2=(1-\varepsilon)^{-1}$ where $\varepsilon=1/6$
and $\varepsilon=0.5$, respectively. We have chosen $n=m=1$. Our results are given in Fig.\ref{fig4}.
Interestingly, we find that the branch of semilocal--dark string solutions
bifurcates with the branch of embedded Abelian--Higgs--dark string solutions
exactly at $\beta_1=\beta_2=(1-\varepsilon)^{-1}$. For $\beta_1 > (1-\varepsilon)^{-1}$, the energy per unit length
of the semilocal--dark string solutions is always smaller than that of the corresponding
embedded Abelian--Higgs--dark string solutions, for $\beta_1 < (1-\varepsilon)^{-1}$ no semilocal--dark string
solutions exist at all. Hence, we find that 
\begin{equation}
 \beta_1^{cr}(\beta_2=(1-\varepsilon)^{-1},1,1)=(1-\varepsilon)^{-1}
\end{equation}

We have also studied the dependence of $\beta_1^{cr}$ on the winding of the dark string
and the Higgs to gauge boson ratio $\beta_2$ of the U(1) model describing the dark string in more detail.
Our results are shown in Fig.\ref{fig3}. Obviously, $\beta_1^{cr}$ increases with increasing
$\varepsilon$. This is related to the fact that the core width of the strings decreases with increasing
$\varepsilon$. This means more gradient energy and hence we have to choose larger values
of $\beta_1$ to be able to compensate for this increase by decrease in potential energy.
 
For $\beta_1=1.0$, which in fact corresponds to the BPS limit of the U(1) dark string model 
for $\varepsilon=0$, the value of $\beta_1^{cr}$ increases for increasing winding $m$ of the
dark string. Again increasing $m$ increases gradient energy such that we have to choose
larger value of $\beta_1$ to compensate the increase by decrease in potential energy.
This is also true when increasing $\beta_2$. Increasing $\beta_2$ decreases the core size of
the dark string, this increases gradient energy and we again have to compensate by increasing
the value of $\beta_1$.

\section{Conclusions}
In this paper we have shown that the interaction of semilocal strings with dark strings
has important effects on the properties of the former. While embedded Abelian--Higgs strings
exist for all values of the Higgs to gauge boson ratio when interacting with dark strings, semilocal
strings with a condensate inside their core exist only above a critical value of the
Higgs to gauge boson ratio. At this critical value, the embedded
Abelian--Higgs--dark strings become unstable. The critical value of the ratio 
depends on the choice of the Higgs to
gauge boson ratio of the dark string and the windings. In the limit where the
ratio tends to the critical ratio, the condensate vanishes identically and the branch of
semilocal--dark string solutions bifurcates with the branch of embedded Abelian--Higgs--dark string
solutions. Apparently, the presence of the condensate lowers the energy in such a way
that whenever semilocal--dark strings exist, they are lower in energy than their
embedded Abelian--Higgs-dark string counterparts. The value of the Higgs to gauge boson ratio
for which semilocal--dark strings exist depends on the value of the condensate
on the string axis and increases for increasing values of the condensate.
All these results are quite different from what is observed in the non--interacting case.
In the non--interacting case, semilocal strings exist only for Higgs to gauge boson
ratio equal to unity and in this limit, the energy per unit length is independent of the value
of the condensate and in addition fulfills a BPS bound. To state it differently~: when not interacting
with dark strings, semilocal strings and embedded Abelian--Higgs strings are degenerate
in energy, while the former are lower in energy as soon as they interact with dark strings.
Since the branch of semilocal--dark string solutions bifurcates with the branch of
embedded Abelian--Higgs--dark strings at the self--dual point of the embedded Abelian--Higgs-dark strings
-- at which these fulfill an energy bound \cite{ha} --  we expect that semilocal--dark strings
are stable. Moreover, they are stable for all choices of the Higgs to gauge boson ratio for which they exist and not just -- as in the non--interacting case -- for Higgs to gauge boson ratio smaller
or equal to unity. Since all current observations point to the fact that the Higgs boson
mass is larger than the gauge boson masses, semilocal strings could still be stable
when interacting with dark strings.
Interestingly -- as mentioned above -- semilocal strings can lower their energy by 
forming a non--vanishing condensate inside their core. This could be important
for the evolution of cosmic string networks since next to the formation of bound states \cite{wyman}
this would be a further mechanism for the network to loose energy.

We didn't study the gravitational properties of the solutions since we believe that the qualitative
features are similar to the case studied in \cite{ha}. Since semilocal--dark strings
have lower energy per unit length than their embedded Ablian--Higgs--dark string
counterparts, we would expect the deficit angle created by the former to be smaller than that of the latter. Furthermore, the critical value of the gravitational coupling at which the solutions
become singular is larger for the semilocal--dark strings than for the
embedded Abelian--Higgs--dark strings.

\end{document}